\documentclass[reprint,twocolumn,superscriptaddress,showpacs,showkeys,preprintnum bers,amsmath,amssymb,floatfix,aps,prl]{revtex4-1}

\usepackage{graphicx}
\usepackage{bm}

\usepackage{placeins}
\usepackage{graphics}
\usepackage{color}
\usepackage{hyperref}
\usepackage{multirow}
\usepackage{blindtext}

\usepackage{pdfpages}

\makeatletter
\AtBeginDocument{\let\LS@rot\@undefined}
\makeatother

\begin{document}


\title{Swapping Exchange and Spin-Orbit Coupling in 2D van der Waals Heterostructures}

\author{Klaus Zollner}
	\email{klaus.zollner@physik.uni-regensburg.de}
	\affiliation{Institute for Theoretical Physics, University of Regensburg, 93053 Regensburg, Germany}	
\author{Martin Gmitra}
	\affiliation{Institute of Physics, P. J. \v{S}af\'{a}rik University in Ko\v{s}ice, 04001 Ko\v{s}ice, Slovakia}
\author{Jaroslav Fabian}
	\affiliation{Institute for Theoretical Physics, University of Regensburg, 93053 Regensburg, Germany}
\date{\today}

\begin{abstract}
The concept of swapping  the two most important spin interactions---exchange and spin-orbit coupling---is proposed based on two-dimensional multilayer van der Waals heterostructures. Specifically, we show by performing realistic \emph{ab initio} simulations, that a \textit{single} device consisting of a bilayer graphene sandwiched 
by a 2D ferromagnet Cr$_2$Ge$_2$Te$_6$ (CGT) and a monolayer WS$_2$, is able not only to  generate, but also to swap the two interactions. The highly efficient swapping is enabled by the interplay of gate-dependent layer polarization in bilayer graphene and short-range spin-orbit and exchange proximity effects affecting only the layers in contact with the sandwiching materials.  We call these structures \textit{ex-so-tic}, for supplying either 
exchange (ex) or spin-orbit (so) coupling in a single device, by gating. Such \textit{bifunctional} devices demonstrate the potential of van der Waals spintronics 
engineering using 2D crystal multilayers.

\end{abstract}

\maketitle

\paragraph{Introduction.}

Novel 2D van der Waals heterostructures give a strong push to spintronics \cite{Zutic2004:RMP,Han2014:NN, Avsar2019:arxiv} which requires 
electrical control of spin interactions to realize devices such as spin transistors
\cite{Datta1990:APL,Schliemann2003:PRL, Wunderlich2010:S,Betthausen2012:S,Chuang2015:NN}.
The past years have seen impressive progress in tuning
spin-orbit and exchange couplings individually \cite{Wang2015:NC, Avsar2017:ACS, Luo2017:NL, Zhong2020:NN, Karpiak2019:arxiv, Ghiasi2017:NL, Ghiasi2019:NL,Ghazaryan2018:NE,Safeer2019:NL,Zutic2019:MT, Benitez2020:NM,Cortes2019:PRL, Zihlmann2018:PRB}. A particularly suitable platform for gating proximity effects
is based on bilayer graphene (BLG). It was proposed that
spin-orbit coupling in BLG can be turned on and off on demand \cite{Gmitra2017:PRL,Khoo2017:NL}, as recently demonstrated \cite{Island2019:Nat}. 
Similar effects have been predicted for ferromagnetic encapsulation \cite{Michetti2010:NL, Cardoso2018:PRL, Zollner2018:NJP}.
The next milestone would be a transport demonstration with sensitivity to the spin polarization of the bands, or the spin-orbit torque in the proximity setup \cite{Alghamdi2019:NL,Kapildeb2019:arxiv, MacNeill2017:NP}.

\begin{figure}[!htb]
	\includegraphics[width=0.97\columnwidth]{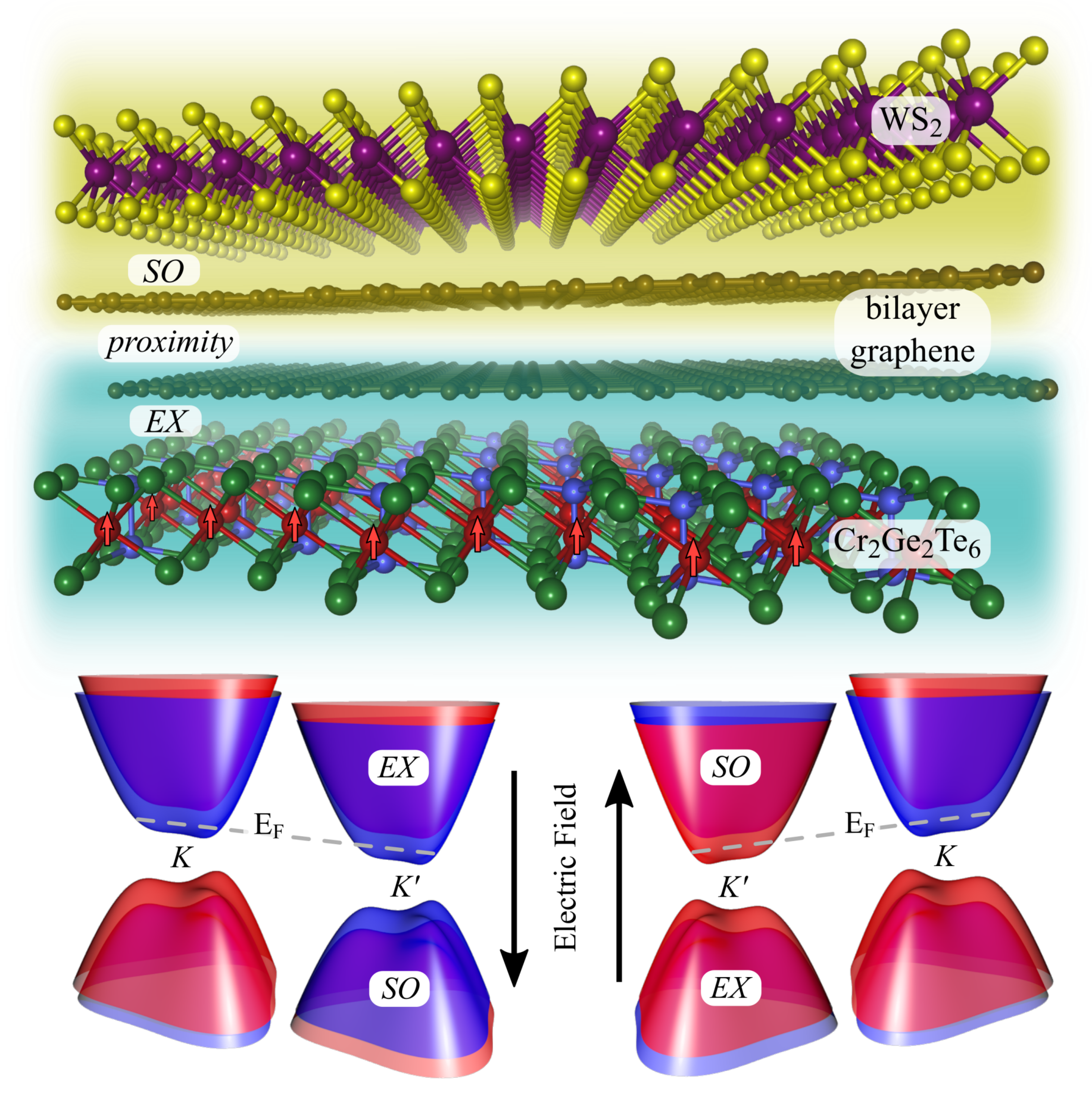}
	\caption{Ex-so-tic van der Waals heterostructure. Top: BLG sandwiched between a monolayer TMDC (such as WS$_2$ or MoSe$_2$) and a monolayer ferromagnetic semiconductor (such as CGT or CrI$_3$). The magnetization of the ferromagnet is indicated by the arrows. The top layer of BLG is proximitized by the TMDC, acquiring a giant spin-orbit coupling, while the bottom layer of the BLG is proximitized by the ferromagnet, acquiring an exchange coupling. Bottom: Electric tunability of the low-energy bands at $K$ and $K'$. 
	The colors red and blue indicate out-of-plane spin. For a fixed Fermi level $E_F$ (here in the conduction band), the Dirac electrons experience either exchange or spin-orbit coupling, depending on the electric field. 
 \label{Fig:device}}
\end{figure}

In this Letter we present a device structure which is capable of not only creating the two most important spin interactions---spin-orbit and exchange couplings---in an electronic system, but also swapping them on demand by an applied electric field. Being able to swap two different spin (or other effective) interactions by gating is a striking thought, without precedence in the realm of conventional materials.  While proximity spin-orbit coupling preserves time-reversal symmetry, and leads to such phenomena as topological quantum spin Hall effect \cite{Kane2005:PRL} or giant spin relaxation anisotropy \cite{Cummings2017:PRL, Omar2019:PRB, Benitez2018:NP, Ghiasi2017:NL}, proximity exchange coupling breaks time reversal symmetry and renders a nominally nonmagnetic electronic system effectively magnetic \cite{Yang2013:PRL}. Swapping the two interactions provides a reversible route between time-reversal symmetric and magnetic physics. 

Our choice of the electronic platform for swapping the spin interactions is BLG, which is nicely suited for proximity-based devices by providing two coupled surfaces. Via layer polarization, which locks a given layer to 
an electronic band (or set of bands), transport properties can be strongly influenced by the environment. We use a CGT/BLG/WS$_2$ heterostructure, with monoloayer Cr$_2$Ge$_2$Te$_6$ (CGT) providing a strong proximity exchange effect to the bottom layer of BLG, and
WS$_2$ to induce strong spin-orbit coupling to the top layer, see Fig.~\ref{Fig:device}. 

Gating can swap spin-orbit and exchange couplings, as illustrated in Fig.~\ref{Fig:device}. The applied electric
field changes the layer polarization, which changes the layer-band assignment, thereby swapping the proximity exchange and spin-orbit couplings. This intuitive
picture is supported below by realistic density functional theory (DFT) simulations and phenomenological modeling, predicting quantitatively the behavior of the ex-so-tic heterostructures in the presence of a transverse electric field.

\paragraph{Swapping spin-orbit and exchange coupling by gate.}

\begin{figure}[!htb]
	\includegraphics[width=0.9\columnwidth]{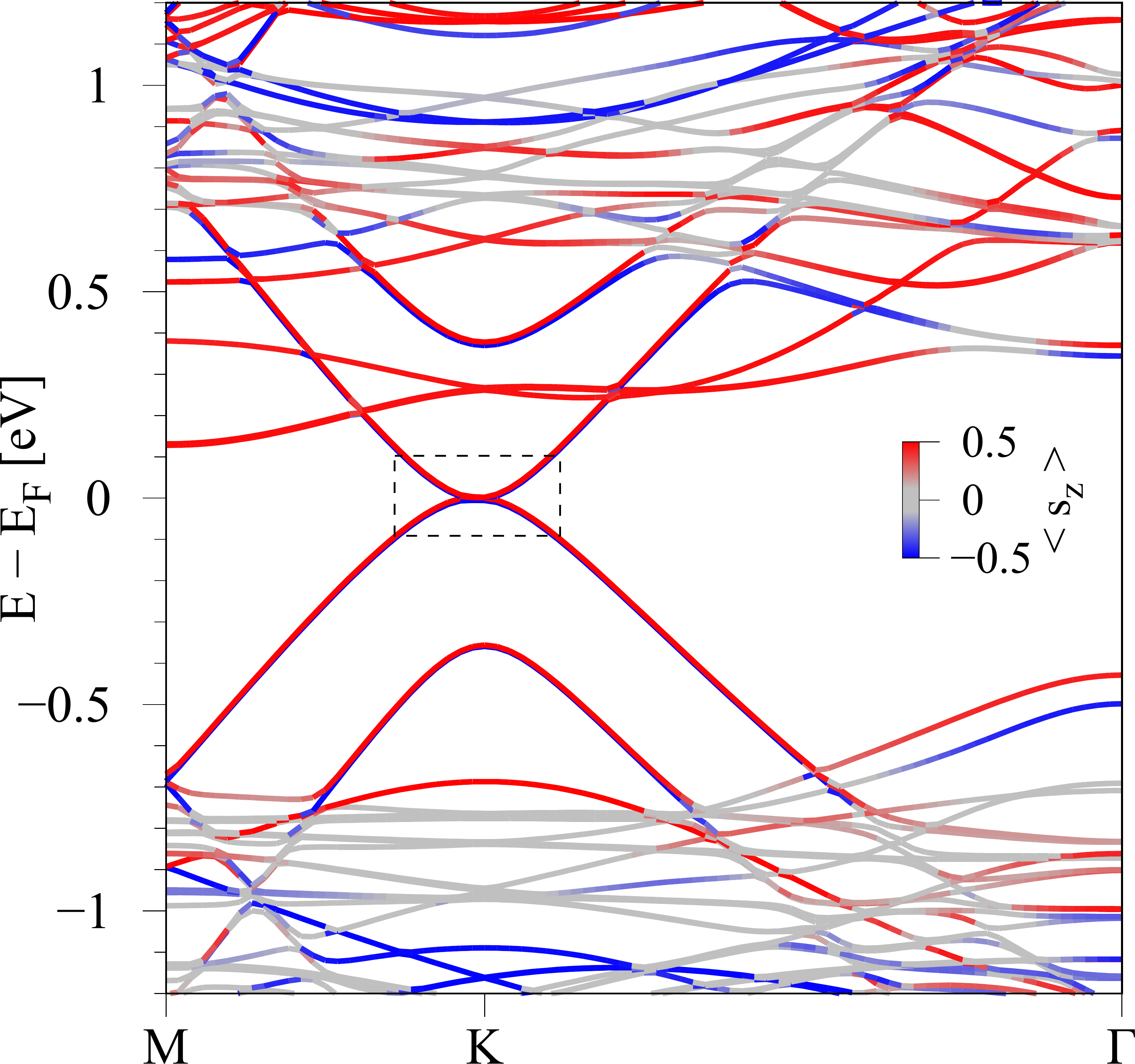}
	\caption{Calculated electronic band structure 
	of a WS$_2$/BLG/CGT stack along the high symmetry path M-K-$\Gamma$. The color of bands (red and blue) corresponds to the $s_z$ spin expectation value. We focus on the low energy bands within the dashed-line box. 
 \label{Fig:bands}}
\end{figure}

We consider a supercell stack containing BLG sandwiched between a monolayer
WS$_2$ and a monolayer ferromagnetic CGT  with an out-of-plane magnetization, as depicted in Fig.~\ref{Fig:device}. We calculate the electronic states for this structure using DFT, see Supplemental Material \footnotemark[1]. The
band structure along selected high-symmetry lines containing the $K$ point is shown in Fig.~\ref{Fig:bands}.
The magnetization --- from CGT --- as well as the strong spin-orbit coupling --- from the transition-metal dichalcogenide (TMDC) --- are manifested by the spin polarization of the bands. Inside the semiconductor band gap, there are well preserved parabolic electronic states of BLG. States relevant for transport form the four low-energy bands close to the Fermi level. 
It is these four bands that are at the focus of this work.

BLG {\it per se} has no band gap, has tiny spin-orbit coupling (on the order of tens of $\mu$eVs \cite{Konschuh2012:PRB}), and has no magnetic exchange coupling. But sandwiched by the two monolayers, both orbital and spin properties of the low-energy bands of BLG strikingly change. First, the built-in dipole moment in the heterostructure \cite{Gmitra2017:PRL, Zollner2018:NJP}
separates the conduction and valence bands, inducing a band gap for a given spin polarization. Next, the states exhibit a giant spin-orbit coupling induced from the TMDC layer. This coupling is opposite in $K$ and $K'$ points, as a consequence of time reversal symmetry of the spin-orbit interaction. Finally, the electronic states of BLG become magnetic, manifested by an exchange splitting, equal in $K$ and $K'$, coming from the ferromagnetic CGT. It is the fascinating spectral separation of the spin-orbit coupling and exchange in the BLG which allows for their swapping. 

\begin{figure*}[htb]
	\includegraphics[width=0.99\textwidth]{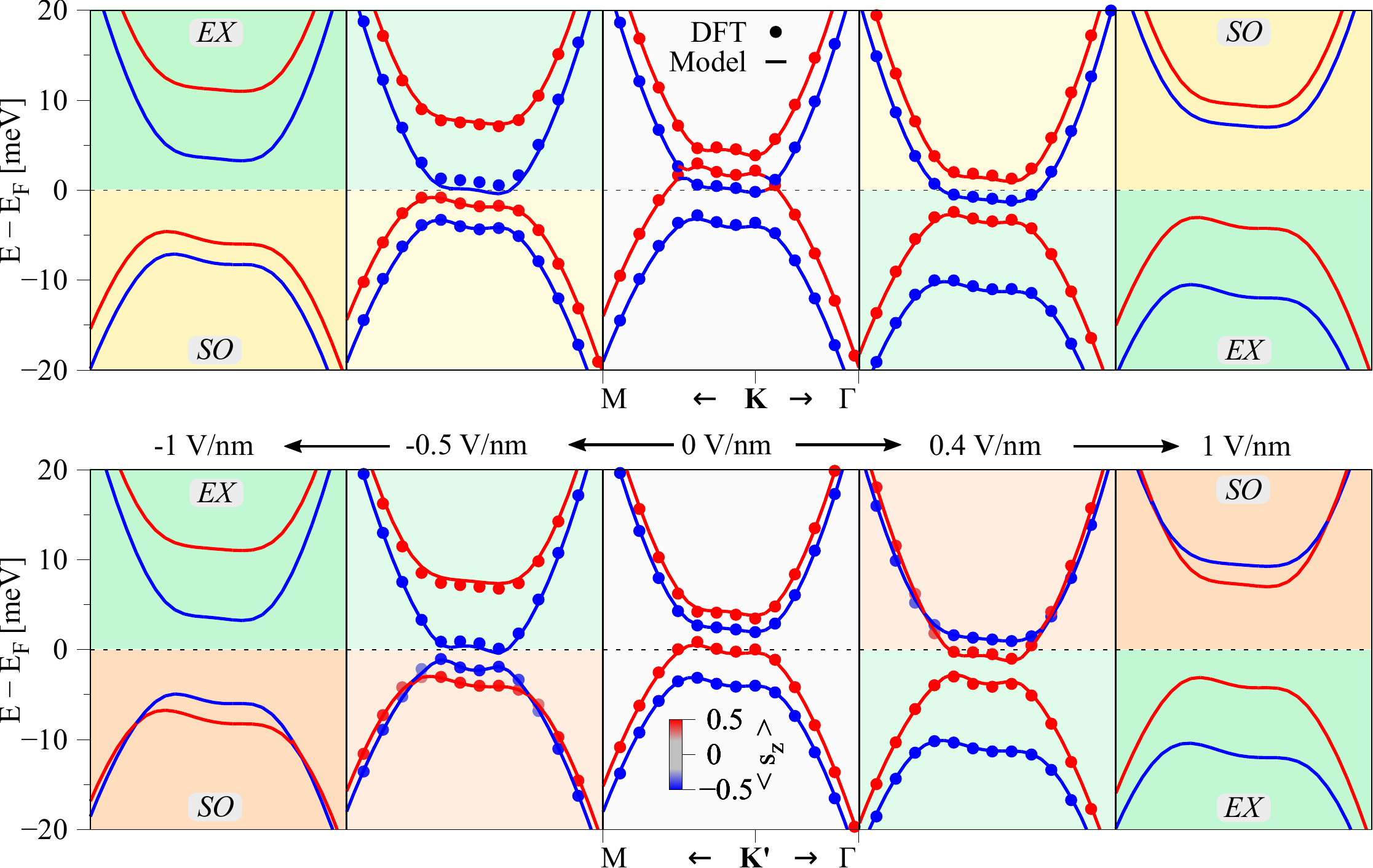}
	\caption{Low energy dispersion curves around K (top) and K' (bottom) for the WS$_2$/BLG/CGT heterostructure. The color of the curves corresponds to the $s_z$ spin expectation value: spin up is red and spin down is blue. The yellowish and reddish backgrounds indicate that the bands are split by spin-orbit coupling (SO): the yellowish is for down-up (blue-red) and reddish for up-down (red-blue) spin ordering along increasing energy. The greenish background is for exchange coupling (EX) whose ordering is always up-down, 
	fixed by the magnetization of the CGT layer.   
	From left to right the transverse electric field is tuned from $-1$ to $1$ V/nm. For electric fields of $-0.5$, $0$, and $0.4$~V/nm both  model (solid lines) and DFT data (symbols) are plotted. For electric fields of $\pm 1$~V/nm we use the model parameters with extrapolated values for $V$, assuming a linear dependence on the field.
	Parameters $V$ = 7.1, 3.2, $-0.5$, $-3.6$, $-8.1$~meV correspond to the field values of 
	$-1$, $-0.5$, 0, 0.4, 1~V/nm, respectively.
 \label{Fig:Efield}}
\end{figure*}

The three aforementioned effects are nicely seen in 
Fig.~\ref{Fig:Efield} which shows a zoom to the low-energy band structure of the stack at both $K$ and $K'$. Perhaps the most striking feature of the band structure is the difference in the energy dispersion at $K$ and $K'$, which ultimately comes from the interplay between the induced spin-orbit and exchange couplings in BLG. 
While the valence and conduction bands partially overlap at $K$, there is a local band gap at $K'$. The dipole moment of the heterostructure is too weak to open a global (over the whole Brillouin zone) band gap. The spin splittings are 2--8~meV, which is experimentally significant.

An applied external electric field increases the band gap of the doubly proximitized BLG and makes explicit the 
effects of the TMDC and CGT layers. This is demonstrated in Fig.~\ref{Fig:Efield}, 
which shows both the DFT results (for fields 
$-0.5$, 0, and 0.4~V/nm) and model calculations (for fields, $-1$, $-0.5$, 0, 0.4, and 1~V/nm) using an effective Hamiltonian introduced in the next section. Let us first look at negative electric fields, say  $-1$~V/nm, which point down, towards CGT. The conduction bands are split
by about 8~meV, having identical spin polarizations at $K$ and $K'$, which means
that this splitting is due to exchange coupling: the conduction bands are affected by the ferromagnetic CGT and conduction electrons will exhibit transport properties of magnetic conductors. On the other hand, the valence bands are split less, by about 2~meV. More important, the spin polarizations of the two bands are opposite at $K$ and $K'$, signaling time-reversal symmetry: the valence bands experience spin-orbit coupling from WS$_2$. This makes sense. At negative electric fields electrons in the upper layer of BLG have a lower energy, and form the valence band. These electrons are affected by the TMDC layer, which gives them the strong spin-orbit coupling character. We will see in the next section that the spin-orbit coupling is of the valley Zeeman type, which is characteristic for graphene proximitized by a TMDC. Similarly, electron orbitals in the lower layer of BLG have a higher energy, forming the conduction band, which is magnetic due to the presence of CGT. 
If the applied electric field points up, towards the TMDC layer, the situation is reversed and the spin characters of the valence and conduction bands are swapped, see Fig.~\ref{Fig:Efield}.
 
We have thus demonstrated the swapping of exchange and spin-orbit couplings: at a fixed chemical potential (doping level), the  electrons exhibit either exchange
or spin-orbit coupling, not both.
Which coupling is realized depends on the gate. The same would work 
with few-layer TMDCs and ferromagnetic semiconductors encapsulating BLG, since the 
proximity effect is sensitive to the interfacial layers only \cite{Zollner2019a:PRB}.

\paragraph{Model Hamiltonian.}

\begin{table*}[htb]
\caption{\label{tab:fitresult} Parameters of the model Hamiltonian $\mathcal{H}$,
Eq.~\eqref{eq:Hamiltonian}, fitted to the DFT low-energy dispersion data for the CGT/BLG/WS$_2$ heterostructure at zero applied electric field. Parameters $\gamma$ are in eV, others in meV. 
The dipole is given in debye. Unspecified model parameters are zero. Based on the fit
parameters for the individual BLG/CGT and WS$_2$/BLG subsystems given in the Supplementary Information, 
we assumed $\lambda_{\textrm{I}}^\textrm{A2}=-\lambda_{\textrm{I}}^\textrm{B2}$ and 
$\lambda_{\textrm{ex}}^\textrm{A1}=\lambda_{\textrm{ex}}^\textrm{B1}$ here for the fit. }
\begin{ruledtabular}
\begin{tabular}{c c c c c c c c c c c c}
$\gamma_0$ & $\gamma_1$ & $\gamma_3$ & $\gamma_4$ & $V$ & $\Delta$ & $\lambda_{\textrm{I}}^\textrm{A2}$ & $\lambda_{\textrm{I}}^\textrm{B2}$ & $\lambda_{\textrm{ex}}^\textrm{A1}$ & $\lambda_{\textrm{ex}}^\textrm{B1}$ & $E_D$ & dipole \\
\hline 
2.432 & 0.365 & -0.273 & -0.164 & -0.474 & 8.854 & 1.132 & -1.132 & -3.874 & -3.874 & 0.348 & 0.398 \\
\end{tabular}
\end{ruledtabular}
\end{table*}

To describe the low-energy states of our heterostructure we employ a Hamiltonian for BLG in the presence of a transverse electric field 
\cite{Konschuh2012:PRB}, 
\begin{equation}
\label{eq:Hamiltonian}
 \mathcal{H} = \mathcal{H}_{\textrm{orb}} + \mathcal{H}_{\textrm{soc}}+\mathcal{H}_{\textrm{R}}+\mathcal{H}_{\textrm{ex}}+E_D.
\end{equation}
It comprises orbital (orb), intrinsic spin-orbit (soc), Rashba (R), and  exchange (ex) terms. We also include the Dirac-point energy $E_D$ shift. Below we specify the individual Hamiltonian terms using pseudospin notation. A matrix form of $\mathcal{H}$ is given in the Supplemental Material \footnotemark[1].

The orbital physics is captured by 
\begin{align}
  \mathcal{H}_{\textrm{orb}} = & -\frac{\sqrt{3}\gamma_0a}{2}\mu_0\otimes(\tau k_x \sigma_x+k_y\sigma_y)\otimes s_0\nonumber\\
  & +\frac{\gamma_1}{2}(\mu_x\otimes\sigma_x-\mu_y\otimes\sigma_y)\otimes s_0\nonumber\\
 & -\frac{\sqrt{3}\gamma_3a}{4}\mu_x\otimes (\tau k_x \sigma_x-k_y\sigma_y)\otimes s_0\nonumber\\
 & -\frac{\sqrt{3}\gamma_3a}{4}\mu_y\otimes(\tau k_x \sigma_y+k_y\sigma_x)\otimes s_0\nonumber\\
  & -\frac{\sqrt{3}\gamma_4a}{2}(\tau k_x \mu_x-k_y\mu_y)\otimes\sigma_0\otimes s_0\nonumber\\
  &+V\mu_z\otimes\sigma_0\otimes s_0\nonumber\\
 & +\Delta(\mu_{+}\otimes\sigma_{+}+\mu_{-}\otimes\sigma_{-})\otimes s_0,
\end{align}
where we denote the graphene lattice constant $a$ and the wave vectors $k_x$ and $k_y$, measured from $\pm$K for the valley index $\tau = \pm 1$. 
The Pauli matrices $\mu_i$, $\sigma_i$, and $s_i$, represent layer, pseudospin, and spin, 
with $i = \{ 0,x,y,z \}$. We also define $\mu_{\pm} = \frac{1}{2}(\mu_z\pm \mu_0)$ and
$\sigma_{\pm} = \frac{1}{2}(\sigma_z\pm \sigma_0)$ to shorten  notation.
Parameters $\gamma_j$, $j = \{ 0,1,3,4 \}$, denote intra- and interlayer hoppings of the BLG. Transverse displacement field 
is introduced by voltage $V$ for the lower, and $-V$ for the upper layer of BLG. Finally, $\Delta$ is the asymmetry in the energy shift of the bonding and antibonding states.

The intrinsic spin-orbit coupling term, while also present in a free-standing BLG, is strongly renormalized by the proximity to the TMDC. This
effect is described by
 \begin{align}
 \mathcal{H}_{\textrm{soc}} =& ~\mu_{+}\otimes \tau(\lambda_{\textrm{I}}^\textrm{A1}\sigma_{+} + \lambda_{\textrm{I}}^\textrm{B1}\sigma_{-}) \otimes s_z\nonumber\\
  &-\mu_{-}\otimes \tau (\lambda_{\textrm{I}}^\textrm{A2}\sigma_{+} + \lambda_{\textrm{I}}^\textrm{B2}\sigma_{-}) \otimes s_z,
  \end{align}
with parameters $\lambda_{\textrm{I}}$ denoting the proximity spin-orbit coupling of the corresponding layer ($1, 2$) and sublattice (A, B) atom. Because of the short-rangeness of the proximity effect only the upper layer is affected, so that only parameters 
$\lambda_{\textrm{I}}^\textrm{A2}$ and $\lambda_{\textrm{I}}^\textrm{B2}$
are significant (on the meV scale). The Rashba coupling can emerge due to the breaking of the space inversion symmetry in the heterostructure and the applied electric field. This term has the form
  \begin{equation}
 \mathcal{H}_{\textrm{R}} = \frac{1}{2}(\lambda_0\mu_z+2\lambda_{\textrm{R}}\mu_0)\otimes(\tau\sigma_x\otimes s_y-\sigma_y\otimes s_x), 
  \end{equation}
where $\lambda_0$ describes the local (intrinsic) breaking of space inversion due to the presence of the other layer in BLG. The resulting spin-orbit fields are opposite in the two layers, giving no net effect on the spin-orbit splitting. The global breaking of space inversion due to the heterostructure and the electric field is accounted for by the proper Rashba parameter $\lambda_{\textrm{R}}$. For a more detailed description of the model and parameters, we refer the reader to Ref. \cite{Konschuh2012:PRB}.
Finally, the magnetic proximity effect induces exchange coupling in BLG, which has the standard form,
\begin{align}
 \mathcal{H}_{\textrm{ex}} =& ~\mu_{+}\otimes (-\lambda_{\textrm{ex}}^\textrm{A1}\sigma_{+} + \lambda_{\textrm{ex}}^\textrm{B1}\sigma_{-}) \otimes s_z \nonumber\\
 & -\mu_{-}\otimes  (-\lambda_{\textrm{ex}}^\textrm{A2}\sigma_{+} + \lambda_{\textrm{ex}}^\textrm{B2}\sigma_{-}) \otimes s_z,
 \end{align}
where parameters $\lambda_{\textrm{ex}}$ represent the proximity exchange for the individual sublattices and layers. 

The spectrum of the effective Hamiltonian $\mathcal H$ is fitted
to the DFT-obtained low-energy dispersion of doubly proximitized BLG at zero applied electric field shown in
Fig.~\ref{Fig:bands}. First, the orbital parameters
$\gamma_j$ from ${\mathcal H}_{\rm orb}$ are obtained, as well as the built-in bias $V$, staggered potential $\Delta$, and the Dirac energy $E_D$. 
In the second step we analyze the fine structure of the spectrum 
and extract proximity induced spin-orbit and exchange parameters. The induced spin-orbit coupling in the upper layer is of the valley-Zeeman type, with opposite parameters
on the two atoms of the sublattice:  
$\lambda_{\textrm{I}}^\textrm{A2} = - \lambda_{\textrm{I}}^\textrm{B2} \approx 1.1$ meV.
The atoms of the lower BLG layer experience strong
proximity exchange coupling due to the adjacent CGT, $\lambda_{\textrm{ex}}^\textrm{A1} = \lambda_{\textrm{ex}}^\textrm{B1} \approx -3.9$ meV .
As for the Rashba coupling, it is negligible (on the
meV scale), which is consistent with our finding that the spin polarization of the considered bands is predominantly out of plane. The fitted parameters
are summarized in Table \ref{tab:fitresult}; parameters not presented there were found to be negligible and set to zero.
With such a minimal set of parameters the agreement
of the effective model and the DFT is excellent, see Fig.~\ref{Fig:Efield}. 

The fitted parameters demonstrate the message that one layer of BLG  experiences giant (on the meV scale) spin-orbit coupling, while the other layer giant exchange coupling, at zero electric field. The only parameter that significantly changes when an electric field is applied is $V$. We therefore change $V$ in our model Hamiltonian (and keep other parameters as fitted to the zero-field DFT results) to see how the spectra develop. The results are presented in Fig.~\ref{Fig:Efield}. Their qualitative interpretation was already given above. Here we only note that the consistency of this procedure (changing only $V$ as the electric field is applied) is checked by performing DFT simulations for two electric fields, $-0.5$ and $0.4$ V/nm; for $\pm 1$~V/nm we use $V$ assuming its linear dependence on the applied electric field. The agreement with the model calculations, seen in Fig.~\ref{Fig:Efield}, gives us full confidence in our approach. More detailed fit results can be found in the Supplemental Material \footnotemark[1], where we also provide results for the individual BLG/CGT and WS$_2$/BLG subsystems, to further validate the robustness of the model. 

For an experimental realization of the swapping effect it is preferred to dope BLG with holes,
to prevent spilling out the conduction-band carriers into the CGT (the Dirac point is
about 100 meV below the conduction band of CGT, see Fig.~\ref{Fig:bands}), upon application of a field. However, we do not expect such a spill-out for conduction-electrons to happen at fields lower than 
0.5 V/nm \cite{Zollner2018:NJP}. 

\paragraph{Summary.}
Doubly proximitized BLG offers a unique platform for investigating fundamental spin physics and designing multifunctional spintronics applications. Using realistic DFT simulations and phenomenological modeling we demonstrate swapping
the two most important spin interactions --- exchange and spin-orbit couplings --- in 
a WS$_2$/BLG/CGT multilayer.
The swapping also means turning the time-reversal symmetry on and off, on demand, in the electronic states at a given doping. Since the effect is robust, we expect a variety
of swapping phenomena 
if the bilayer sandwich comprises antiferromagnets, ferroelectricity, ferroelectrics, topological insulators, or superconductors.

\footnotetext[1]{See Supplemental Material including Refs. \cite{Baskin1955:PR,Kresse1999:PRB,Perdew1996:PRL,Grimme2006:JCC,Barone2009:JCC,Bengtsson1999:PRB, Carteaux1995:JPCM,Gong2017:Nat, Schutte1987:JSSC, Zollner2018:NJP, Gmitra2017:PRL, Hohenberg1964:PRB,Konschuh2012:PRB,Island2019:Nat,Giannozzi2009:JPCM,Wang2019:APL,Chen2015:PRB,Li2014:JMCC}, where we present the computational details and results for the individual BLG/CGT and WS$_2$/BLG heterostructures.
In addition, we show an animation about the electric field evolution of the low energy bands of doubly proximitized BLG. }

\acknowledgments
This work was funded by the Deutsche Forschungsgemeinschaft (DFG, German Research Foundation) SFB 1277 (Project-ID 314695032), the European Unions Horizon 2020 research and innovation program 
under Grant No. 785219, DFG SPP 1666,  
and by the VEGA 1/0105/20 and the Internal Research Grant System VVGS-2019-1227.

\bibliographystyle{naturemag}
\bibliography{paper}

\cleardoublepage


\section*{Supplementary material}

# Supplemental Material:
## Swapping Exchange and Spin-Orbit Coupling in 2D van der Waals Heterostructures

Klaus Zollner,[1, *] Martin Gmitra,[2] and Jaroslav Fabian[1]

[1]*Institute for Theoretical Physics, University of Regensburg, 93053 Regensburg, Germany*
[2]*Institute of Physics, P. J. Šafárik University in Košice, 04001 Košice, Slovakia*

In the Supplemental Material, we show the geometries, band structures, and low energy bilayer graphene (BLG) bands fitted by the model Hamiltonian for the precursor heterostructures $WS_2$/BLG, and BLG/$Cr_2Ge_2Te_6$ (CGT). The results further validate the robustness of the model and accuracy of our fits. Furthermore, we show the exact geometry of the CGT/BLG/$WS_2$ heterostructure from the main text, as well as a more detailed view on the fitted low energy bands for zero and finite electric fields. We also explicitly write the Hamiltonian in matrix representation. In addition, we present details about our structural setup and the first-principles calculations.

## GEOMETRY SETUP

For the considered heterostructure of the main text, we choose a $5 \times 5$ supercell of BLG in Bernal stacking, a $\sqrt{3} \times \sqrt{3}$ CGT supercell and a $4 \times 4$ supercell of $WS_2$. We stretch the lattice constant of graphene by roughly 2% to 2.5 Å [1] and stretch the lattice constant of CGT by roughly 6% to 7.2169 Å [2]. DFT calculations predict that the tensile strain does not change the ferromagnetic ground state of CGT, but the band gap and Curie temperature are enhanced [3–5]. The $WS_2$ lattice constant is compressed by about 1% to 3.125 Å [6]. The supercell of $WS_2$/BLG/CGT has a lattice constant of 12.5 Å and contains 178 atoms in the unit cell. In Fig. S1 we show the exact geometry, akin to Fig. 1 of the main text, including relaxed interlayer distances of the CGT/BLG/$WS_2$ hybrid structure. The obtained interlayer distances are similar as reported in Refs. [7, 8]. The lower (upper) graphene layer, formed by sublattices $A_1$ and $B_1$ ($A_2$ and $B_2$), is proximitized by the CGT ($WS_2$) only, due to the short range proximity effect. Within our assumption of perfectly aligned individual layers, the full heterostructure still has $C_3$ symmetry after relaxation.

## FIRST-PRINCIPLES CALCULATIONS

The electronic structure calculations and structural relaxation of the BLG-based heterostructures are performed by DFT [9] with QUANTUM ESPRESSO [10]. Self-consistent calculations are performed with the $k$-point sampling of $12 \times 12 \times 1$ to get converged results for the proximity exchange and spin-orbit coupling. For the CGT/BLG/$WS_2$ heterostructure, when spin-orbit coupling is included, a smaller $k$-point sampling of $6 \times 6 \times 1$ is used. We perform open shell calculations that provide the spin polarized ground state of CGT. A Hubbard parameter of $U = 1$ eV is used for Cr $d$-orbitals, being in the range of proposed values for CGT [8, 11]. We use an energy cutoff for charge density of 500 Ry, and the kinetic energy cutoff for wavefunctions is 60 Ry for the scalar relativistic pseudopotential with the projector augmented wave method [12] with the Perdew-Burke-Ernzerhof exchange correlation functional [13]. When SOC is included, we use the relativistic versions of the pseudopotentials. For the relaxation of the heterostructures, we add van der Waals corrections [14, 15] and use quasi-newton algorithm based on trust radius procedure. Dipole corrections [16] are also included to get correct band offsets and internal electric fields. In order to simulate quasi-2D systems, we add a vacuum of 20 Å, to avoid interactions between periodic images in our slab geometry. To determine the interlayer distances, the atoms of BLG and $WS_2$ are allowed to relax only in their $z$ positions (vertical to the layers), and the atoms of CGT are allowed to move in all directions, until all components of all forces are reduced below $10^{-3}$ [Ry/$a_0$], where $a_0$ is the Bohr radius.

## HETEROSTRUCTURE OF BLG, CGT, AND WS2

Here we present the Hamiltonian used to model the doubly proximitized BLG. The basis states are $|C_{A1}, \uparrow\rangle$, $|C_{A1}, \downarrow\rangle$, $|C_{B1}, \uparrow\rangle$, $|C_{B1}, \downarrow\rangle$, $|C_{A2}, \uparrow\rangle$, $|C_{A2}, \downarrow\rangle$, $|C_{B2}, \uparrow\rangle$, and $|C_{B2}, \downarrow\rangle$. In this basis the Hamiltonian from the main text is (see also Ref. 17)

$$\mathcal{H} = \mathcal{H}_{\text{orb}} + \mathcal{H}_{\text{soc}} + \mathcal{H}_{\text{ex}} + \mathcal{H}_{\text{R}} + E_D, \tag{S1}$$

$$\mathcal{H}_{\text{orb}} = \begin{pmatrix} \Delta + V & \gamma_0 f(\boldsymbol{k}) & \gamma_4 f^*(\boldsymbol{k}) & \gamma_1 \\ \gamma_0 f^*(\boldsymbol{k}) & V & \gamma_3 f(\boldsymbol{k}) & \gamma_4 f^*(\boldsymbol{k}) \\ \gamma_4 f(\boldsymbol{k}) & \gamma_3 f^*(\boldsymbol{k}) & -V & \gamma_0 f(\boldsymbol{k}) \\ \gamma_1 & \gamma_4 f(\boldsymbol{k}) & \gamma_0 f^*(\boldsymbol{k}) & \Delta - V \end{pmatrix} \otimes s_0, \tag{S2}$$

$$\mathcal{H}_{\text{soc}} + \mathcal{H}_{\text{ex}} + \mathcal{H}_{\text{R}} = \begin{pmatrix} (\tau \lambda_{\text{I}}^{\text{A1}} - \lambda_{\text{ex}}^{\text{A1}}) s_z & \text{i}(\lambda_0 + 2\lambda_{\text{R}}) s_-^\tau & 0 & 0 \\ -\text{i}(\lambda_0 + 2\lambda_{\text{R}}) s_+^\tau & (-\tau \lambda_{\text{I}}^{\text{B1}} - \lambda_{\text{ex}}^{\text{B1}}) s_z & 0 & 0 \\ 0 & 0 & (\tau \lambda_{\text{I}}^{\text{A2}} - \lambda_{\text{ex}}^{\text{A2}}) s_z & -\text{i}(\lambda_0 - 2\lambda_{\text{R}}) s_-^\tau \\ 0 & 0 & \text{i}(\lambda_0 - 2\lambda_{\text{R}}) s_+^\tau & (-\tau \lambda_{\text{I}}^{\text{B2}} - \lambda_{\text{ex}}^{\text{B2}}) s_z \end{pmatrix}. \tag{S3}$$

We use the linearized version for the nearest-neighbor structural function $f(\boldsymbol{k}) = -\frac{\sqrt{3}a}{2}(\tau k_x - \text{i}k_y)$ and introduce also $s_{\pm}^\tau = \frac{1}{2}(s_x \pm \text{i}\tau s_y)$ for shorter notation.

TABLE S1. Fit parameters of the model Hamiltonian $\mathcal{H}$, Eq. (1) in the main text, for the BLG/WS2 structure with SOC, for the CGT/BLG structure without SOC, and for the CGT/BLG/WS2 structure without SOC. Unspecified parameters are zero.

| system | BLG/WS2 (SOC) | CGT/BLG (no SOC) | CGT/BLG/WS2 (no SOC) |
|---|---|---|---|
| $\gamma_0$ [eV] | 2.453 | 2.541 | 2.434 |
| $\gamma_1$ [eV] | 0.372 | 0.384 | 0.365 |
| $\gamma_3$ [eV] | -0.270 | -0.296 | -0.269 |
| $\gamma_4$ [eV] | -0.162 | -0.179 | -0.165 |
| $V$ [meV] | 4.226 | -8.499 | -0.479 |
| $\Delta$ [meV] | 10.208 | 10.434 | 8.649 |
| $\lambda_{\text{I}}^{\text{A2}}$ [meV] | 1.070 | 0 | 0 |
| $\lambda_{\text{I}}^{\text{B2}}$ [meV] | -1.179 | 0 | 0 |
| $\lambda_{\text{ex}}^{\text{A1}}$ [meV] | 0 | -4.567 | -4.257 |
| $\lambda_{\text{ex}}^{\text{B1}}$ [meV] | 0 | -4.224 | -3.997 |
| $E_D$ [meV] | -3.844 | -1.003 | -0.336 |
| dipole [debye] | -1.089 | 1.527 | 0.366 |

In Fig. S1, we present a side view of the CGT/BLG/WS2 heterostructure. The interlayer distance are of typical van der Waals type and similar as obtained in Refs. [7, 8]. In Tab. S1, we summarize the fit results for the CGT/BLG/WS2 stack, when spin-orbit coupling (SOC) is not included in the calculation. The results are similar to the case with SOC, as listed in Tab. (1) in the main text, but SOC parameters are zero.

In Fig. S2 (Fig. S3), we show the calculated low energy bands with the model fit for zero (finite) electric field. Figs. S2 and S3 are more detailed views of the DFT-fitted bands, as shown in Fig. (3) from the main text. In all cases, the model agrees very well with the first-principles data and also band splittings are well reproduced. As described in the main text, we use the DFT-fitted parameters for zero electric field from Tab. (1) and tune $V$ to match the finite electric field cases. An animation (`Efield_movie.mp4`) showing the evolution of the low energy bands of doubly proximitized BLG at $K$ and $K'$, for a series of electric fields is also available.

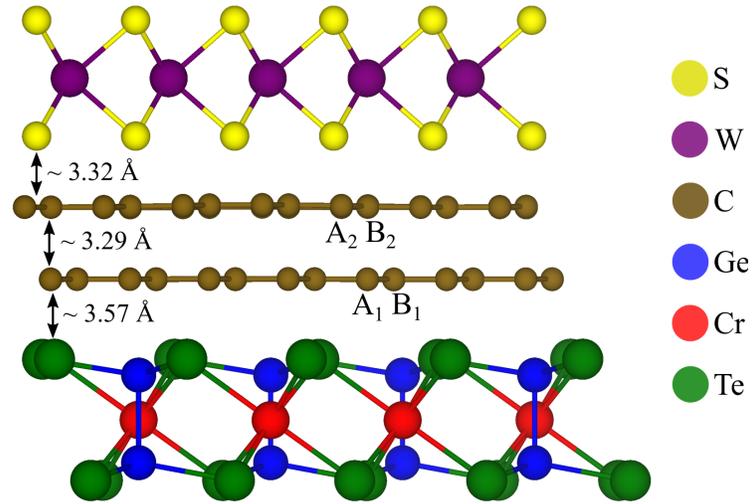

FIG. S1. Side view of the CGT/BLG/WS$_2$ heterostructure. Different colored spheres correspond to the different atomic species. The lower (upper) graphene layer of BLG is formed by sublattices A$_1$ and B$_1$ (A$_2$ and B$_2$). The relaxed interlayer distances are also indicated.

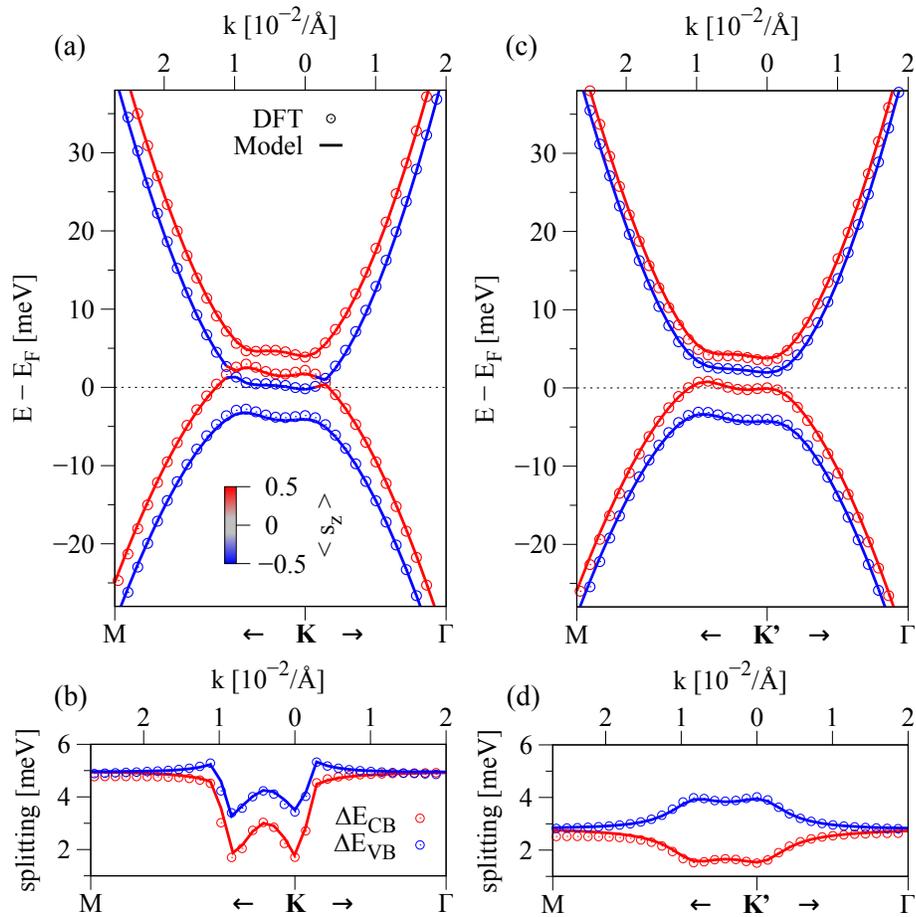

FIG. S2. (a) Zoom to the calculated low energy bands (symbols) around the $K$ point with a fit to the model Hamiltonian (solid lines) for the WS$_2$/BLG/CGT heterostructure including SOC for zero electric field. (b) Energy splittings of the conduction and valence band. (c,d) Same as (a,b), but for the $K'$ point.

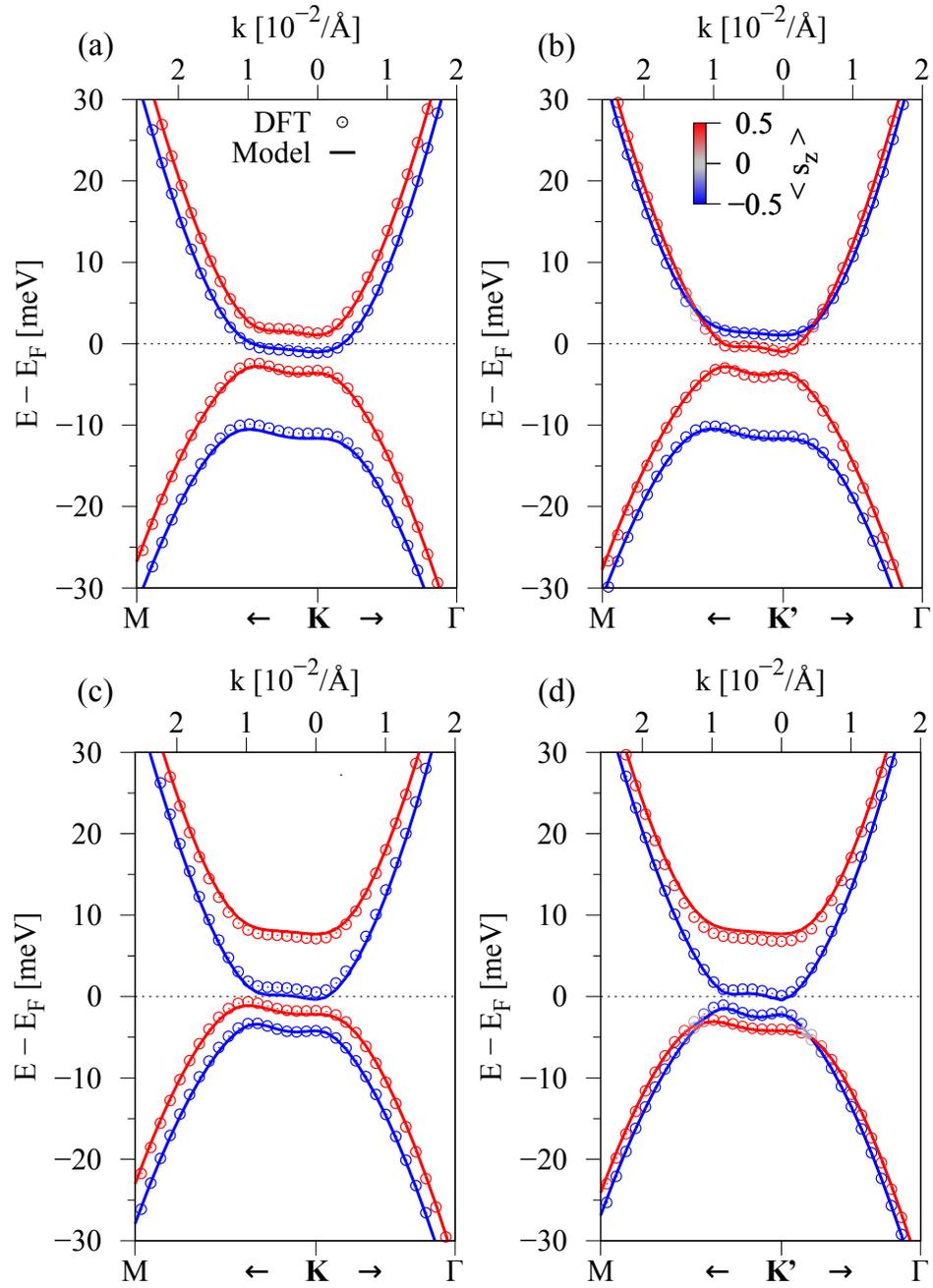

FIG. S3. Calculated (symbols) low energy band structures with a fit to the model Hamiltonian (solid lines), around $K$ and $K'$ points for WS$_2$/BLG/CGT heterostructure including SOC. (a,b) For transverse electric field of $+0.4$ V/nm and a calculated dipole of 2.755 Debye. (c,d) For transverse electric field of $-0.5$ V/nm and a calculated dipole of -2.649 Debye.

## HETEROSTRUCTURE OF BLG AND WS2

For the WS$_2$/BLG heterostructure, we choose a $5 \times 5$ supercell of BLG in Bernal stacking and a $4 \times 4$ supercell of WS$_2$. We stretch the lattice constant of graphene by roughly 2% to 2.5 Å and the WS$_2$ lattice constant is compressed by about 1% from 3.153 Å [6] to 3.125 Å. The supercell of WS$_2$ on BLG has a lattice constant of 12.5 Å and contains 148 atoms in the unit cell.

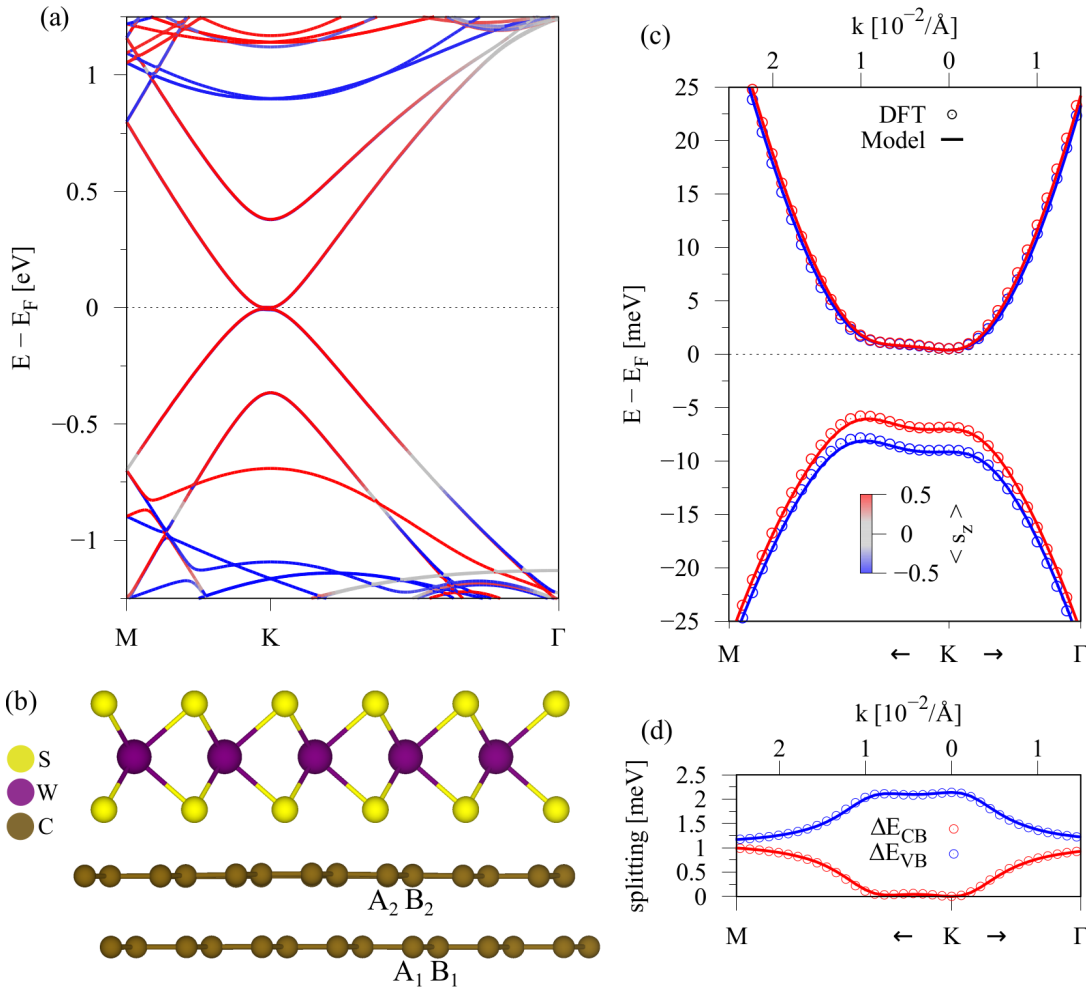

FIG. S4. (a,b) Band structure with SOC and geometry of WS$_2$ on BLG. (c) Zoom to the calculated low energy bands (symbols) around the $K$ point with a fit to the model Hamiltonian (solid lines). The color corresponds to the $s_z$ expectation value. (d) Energy splittings of the conduction and valence band.

Recent first-principles calculations have shown that for BLG on WS$_2$, there exists a spin-valve effect [7]. In Figs. S4(a,b) we show the band structure and a side view of the geometry of WS$_2$ on BLG. The results are in perfect agreement with Ref. [7]. We go one step further and fit our BLG low energy model Hamiltonian from the main text to the DFT calculated bands and find perfect agreement between the model and the ab-initio data, see Figs. S4(c,d). The fit parameters are summarized in Tab. S1 for the BLG/WS$_2$ heterostructure. The low energy valence band originates from the top graphene layer formed by sublattices $A_2$ and $B_2$. Because of the short range nature of proximity SOC, only this top layer is proximitized by the WS$_2$. Therefore the valence band experiences strong SOC induced spin splitting, while the conduction band splitting is much smaller near the $K$ point.

Because of this short rangeness of proximity SOC and the unique BLG band structure, the splitting can be switched from the valence band to the conduction band by applying a transverse electric field across the heterostructure [7, 18], and spin relaxation of electrons and holes can be controlled fully electrically.

## HETEROSTRUCTURE OF BLG AND CGT

For the BLG/CGT heterostructure, we choose a $5 \times 5$ supercell of BLG in Bernal stacking and a $\sqrt{3} \times \sqrt{3}$ CGT supercell. We keep the lattice constant of graphene unchanged at $a = 2.46$ Å and stretch the CGT lattice constant by roughly 4% from 6.8275 Å [2] to 7.1014 Å. The supercell of BLG on CGT has a lattice constant of 12.3 Å and contains 130 atoms in the unit cell.

Similar to before, in Figs. S5(a,b) we show the band structure and a side view of the geometry of BLG on CGT. In this case, the bottom graphene layer formed by sublattices $A_1$ and $B_1$ experiences strong proximity exchange from CGT. This system has recently been considered in Ref. [8], where an exchange valve effect has been found. Again, we fit our low energy model to the first-principles data and find perfect agreement, see Figs. S5(c,d). Here, the valence band, which originates from the bottom graphene layer formed by sublattices $A_1$ and $B_1$, is strongly split due to proximity exchange. In contrast, the conduction band splitting is negligible, again due to short rangeness of the proximity effect. The fit parameters are also summarized in Tab. S1 for the CGT/BLG heterostructure.

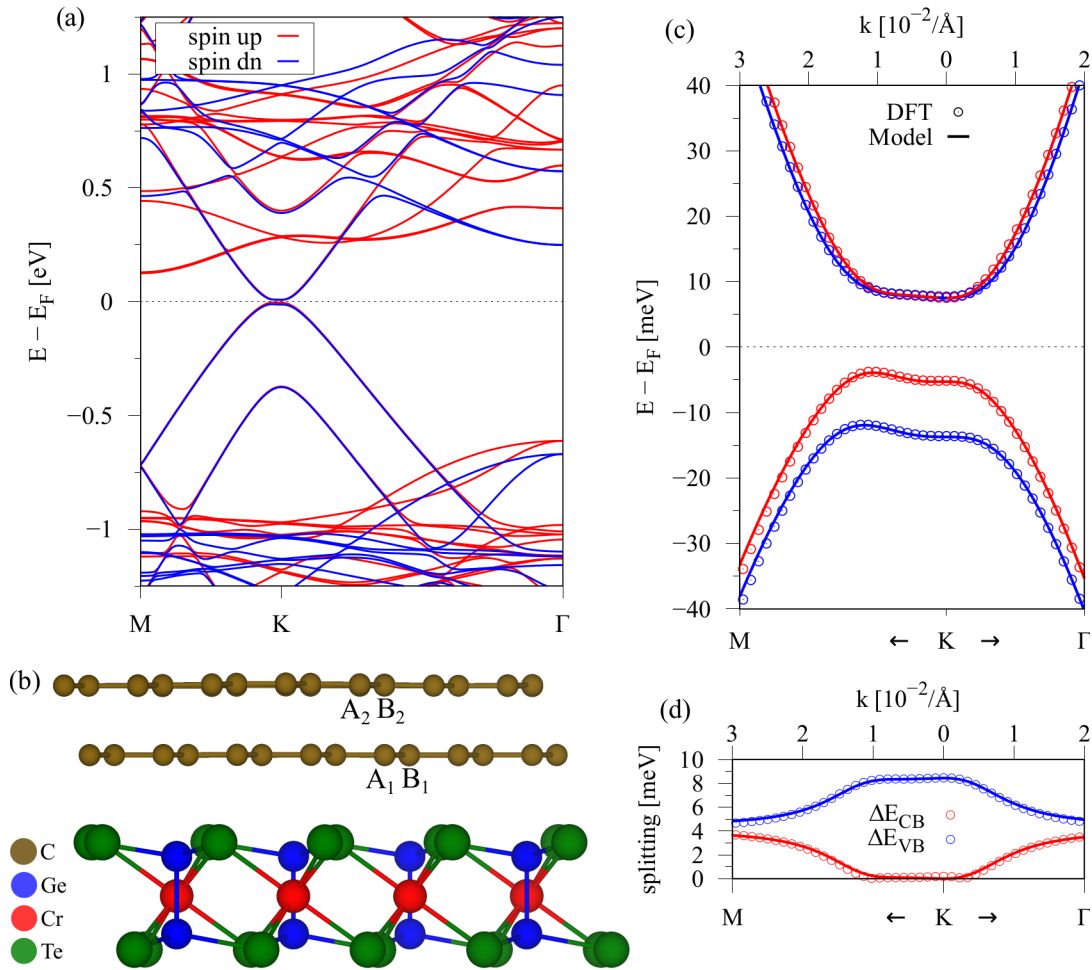

FIG. S5. (a,b) Band structure without SOC and geometry of BLG on CGT. (c) Zoom to the calculated low energy bands (symbols) around the $K$ point with a fit to the model Hamiltonian (solid lines). Bands in red (blue) correspond to spin up (down). (d) Energy splittings of the conduction and valence band.

* klaus.zollner@physik.uni-regensburg.de

\clearpage

\end{document}